\documentclass[%
 reprint,
 longbibliography,
 amsmath,amssymb,
 aps,
 prl,
]{revtex4-2}
\usepackage{graphicx}% Include figure files
\usepackage{dcolumn}% Align table columns on decimal point
\usepackage{bm}% bold math
\usepackage[colorlinks=true,linkcolor=blue,citecolor=red]{hyperref}% add hypertext capabilities
\usepackage{bbm}
\usepackage{physics}
\usepackage{nicefrac}
\usepackage{siunitx}
\usepackage{soul}
\usepackage{color}
\newcommand{\mrm}{\mathrm}
\newcommand{\mcal}{\mathcal}

\newcommand{\pr}[1]{\left(#1\right)}
\newcommand{\sr}[1]{\left[#1\right]}

\newcommand{\kT}{k_{\mathrm{B}}T}
\DeclareMathOperator{\e}{e}

\newcommand{\trel}{\tau_{\mrm{r}}}
\newcommand{\dg}{\delta_{\mrm{g}}}
\newcommand{\ts}{t_{\mrm{s}}}
\newcommand{\fs}{f_{\mrm{s}}}

\newcommand{\wt}{W}
\newcommand{\dtot}{\mrm{d}}
\newcommand{\fc}{f_\mrm{c}}

\newcommand{\xkm}{x_{k-1}}
\newcommand{\xk}{x_{k}}
\newcommand{\xkp}{x_{k+1}}

\newcommand{\ykm}{y_{k-1}}
\newcommand{\yk}{y_{k}}
\newcommand{\ykp}{y_{k+1}}
\newcommand{\Ykm}{Y^{k-1}}
\newcommand{\Yk}{Y^{k}}

\newcommand{\xhk}{\hat{x}_{k}}
\newcommand{\xhkp}{\hat{x}_{k+1}}

\newcommand{\lk}{\lambda_{k}}
\newcommand{\lkp}{\lambda_{k+1}}

\newcommand{\lmd}{\lambda}

\newcommand{\sm}{\sigma_{\mrm m}^{2}}
\newcommand{\ssm}{\sigma_{\mrm m}}

\newcommand{\astar}{\alpha^{*}}
\newcommand{\snr}{\mrm{SNR}}
\newcommand{\snrc}{\mrm{SNR}_{\mrm{c}}}

\newcommand{\beginsupplement}{%
        \setcounter{table}{0}
        \renewcommand{\thetable}{\arabic{table}}%
        \setcounter{figure}{0}
       \renewcommand{\thefigure}{S\arabic{figure}}
       \setcounter{equation}{0}
       \renewcommand{\theequation}{S\arabic{equation}}
        }

\begin{document}

\title{Bayesian information engine that optimally exploits noisy measurements}

\author{Tushar K. Saha}%
\email{tushars@sfu.ca}
\author{Joseph N. E. Lucero}
\altaffiliation{
    Department of Chemistry, Stanford University, Stanford, CA 94305, USA
}
\author{Jannik Ehrich}
\author{David A.\ Sivak}
\email{dsivak@sfu.ca}
\author{John Bechhoefer}
\email{johnb@sfu.ca}
\affiliation{
    Department of Physics, Simon Fraser University, Burnaby, BC, V5A 1S6 Canada
}

\date{\today}

\begin{abstract}

We have experimentally realized an information engine consisting of an optically trapped, heavy bead in water. The device raises the trap center after a favorable ``up'' thermal fluctuation, thereby increasing the bead's average gravitational potential energy. In the presence of measurement noise, poor feedback decisions degrade its performance; below a critical signal-to-noise ratio, the engine shows a phase transition and cannot store any gravitational energy. However, using Bayesian estimates of the bead's position to make feedback decisions can extract gravitational energy at all measurement noise strengths and has maximum performance benefit at the critical signal-to-noise ratio.

% 598 characters without spaces (600 max)
\end{abstract}


\maketitle

Information engines are a new class of engine that use information as fuel to convert heat from a thermal bath into useful energy. They exploit knowledge of thermal fluctuations to apply feedback and extract energy from the thermal bath, while paying costs required by the second law of thermodynamics to process that information~\cite{parrondo2015}. In the past decade, information engines have been realized in a wide range of physical systems~\cite{toyabe2010experimental,camati2016experimental,koski2015chip,chida2017power,kumar2018sorting,thorn2008experimental,vidrighin2016photonic}. For practical application, it is important to understand how to maximize the engine's output~\cite{saha2021trajectory,saha2021maximizing} and efficiency~\cite{paneru2018lossless,admon2018experimental,ribezzi2019large,saha2021maximizing}. 

One obstacle that degrades the output of an information engine is inaccurate information about the system that arises from measurement noise. Since information engines respond to measurements of thermal fluctuations, measurement noise can lead to wrong feedback decisions. Feedback actions chosen based on inaccurate measurements reduce the work extracted from the surrounding thermal bath and can even, at high noise levels, lead to a net heating of the thermal bath~\cite{paneru2020efficiency}. 

Previous efforts to account for noisy measurements in information engines have all used ``naive" feedback algorithms based directly on the most recent noisy measurement~\cite{paneru2020efficiency,taghvaei2022relation}. Here we show that such information engines, with unidirectional ratchets, have a phase transition between working and non-working regimes: Below a critical level of signal-to-noise ratio for measurements of the engine state, a ``pure'' information engine---one that requires no work input beyond that needed to run the measurement and control apparatus---is not possible.

Although previous studies noted the degradation of performance due to measurement noise, they did not attempt to alter the feedback algorithm to compensate. Yet theoretical studies have indicated that incorporating the information contained in past measurements via optimal feedback control could greatly improve the performance of an information engine~\cite{taghvaei2022relation,nakamura2021connection,horowitz2013imitating,rupprecht2020predictive}. Indeed, experiments in other areas of physics have used feedback that incorporates Bayesian estimators to demonstrate spectacular results, even in the presence of high measurement noise; significant achievements include trapping a single fluorescent dye molecule that is freely diffusing in water~\cite{fields2011} and cooling a nanoparticle to the quantum regime of dynamics~\cite{magrini2021real,conangla2019optimal}.

In this Letter, we present an experimental realization of an optimal Bayesian information engine that retains the relevant memory of all past measurements in a single summary statistic. Using the extra information from past measurements and correctly compensating for delays in the feedback loop via predictive estimates, we extract and store significant amounts of energy, even in the presence of high measurement noise. 

Our implementation of the Bayesian filter uses the optimal affine feedback control algorithm~\cite{lucero2021maximal}, at optimal experimental parameters~\cite{saha2021maximizing}, to maximize the engine's rate of gravitational-energy storage. The relevant information from past observations is used to minimize the uncertainty in the bead's position. This Bayesian information engine extracts energy even at low signal-to-noise ratio (SNR), avoiding the phase transition in the naive information engine that leads to zero output. Under any conditions, this engine extracts at least as much work as the naive engine and, indeed, reaches the upper bound on the performance of Gaussian information engines.

\textit{Experimental setup}.---Our information engine consists of a 4-\textmu m heavy bead trapped in a horizontally aligned optical trap. Because of the surrounding heat bath, the bead's position fluctuates about an equilibrium average. The heavy bead is also subject to gravity. The bead's position is measured with a sampling time $\ts=20$~\textmu s using the scattered light from a detection laser.

In the operation of the engine, an observed ``up'' fluctuation increases the bead's gravitational potential energy and can be captured (``rectified'') by a quick feedback response that shifts up the trap position. The response takes place after a one-time-step delay $t_\text{d} = \ts = 20$ \textmu s, with the shift chosen so that the trap does zero work on the bead. For this ``pure'' information engine, the work to run the motor is associated only with the measuring device and feedback controller, not the engine itself.

\begin{figure}[t]
    \centering
    \includegraphics[width=\linewidth]{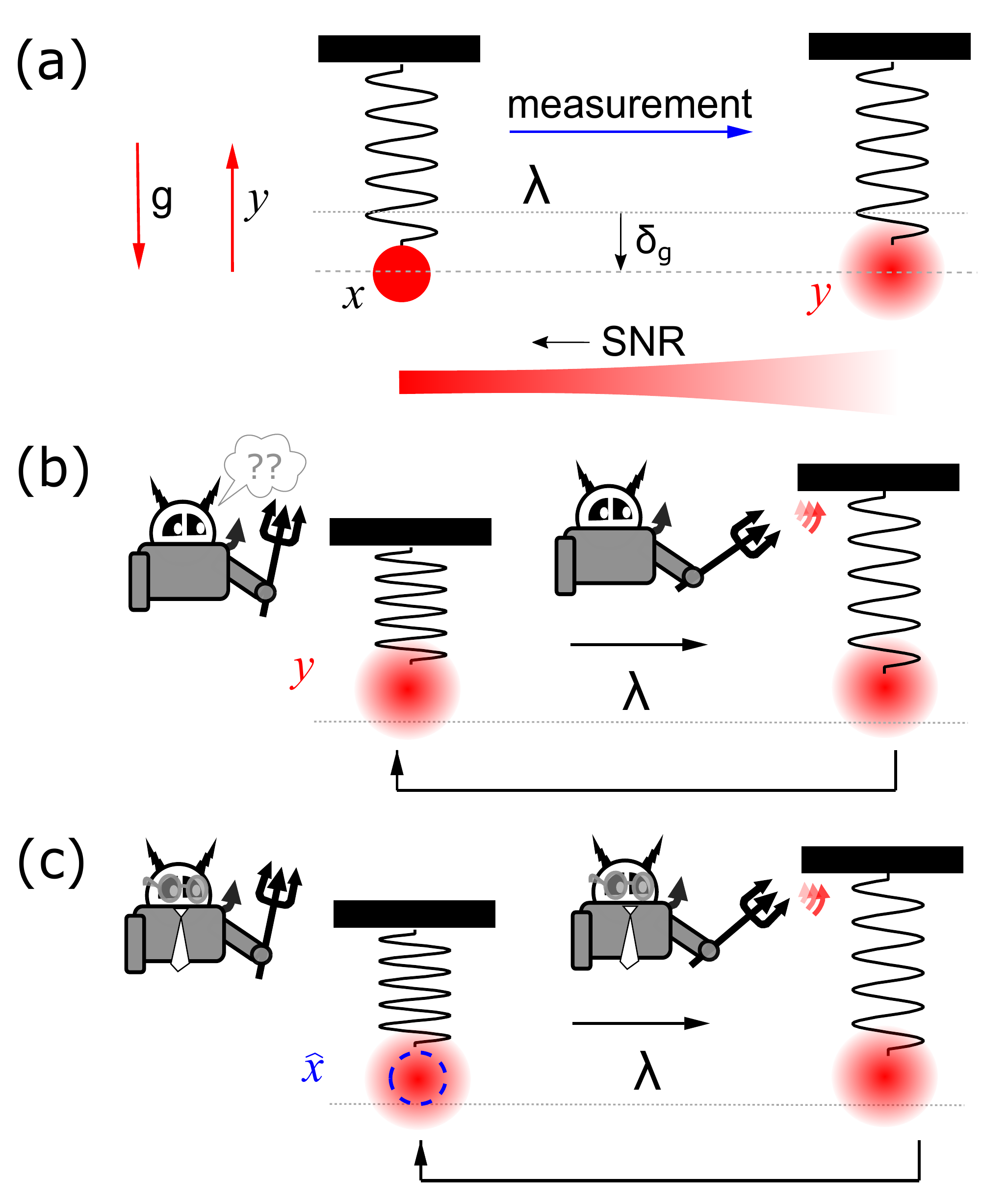}
    \caption{Schematic information engine. (a) Noisy detector measures position $y$ of the bead, actually located at $x$. Ratchet based on either (b) noisy position measurement $y$ or (c) Bayesian position estimate $\hat{x}$ (blue dashed circle).}
    \label{fig:engine_schematic}
\end{figure}

The optically trapped bead can be modeled by a spring-mass system [Fig.~\ref{fig:engine_schematic}(a)]. The true position of the bead is estimated from a noisy measurement $y$. Measurement noise is increased by reducing the intensity of the detection laser beam. 

Figure~\ref{fig:engine_schematic}(b) shows a \textit{naive} information engine that directly uses a noisy measurement $y$ to apply feedback. By contrast, the \emph{Bayesian} information engine bases its feedback on the best estimate $\hat{x}$ of the bead's position [Fig.~\ref{fig:engine_schematic}(c)]. The bead's position is estimated using a Bayesian filter that explicitly models the measurement noise and feedback delay~\footnote{The delay is also termed ``feedback latency,'' which refers explicitly to the sum of delays in the feedback loop, which include contributions from measurements, computations, and output to the actuator (the acousto-optic deflector, which moves the trap center). We use the more familiar informal language in the text.}. 

The information engine can extract energy at high measurement noise because feedback decisions based on filtered estimates of the bead's position are more likely to ratchet to a ``true'' upward fluctuation rather than to measurement noise. 

\textit{Equations of motion}.---The dynamics of the optically trapped bead obey an overdamped Langevin equation,
\begin{align}
    \gamma\dot{x} = -\kappa\sr{x(t) - \lambda(t)} - mg + \sqrt{2\kT\gamma}\ \xi(t) \,,
\end{align}
where $x(t)$ is the position at time $t$ of a bead of diameter $d$ and effective (buoyant) mass $m$ in a trap of stiffness $\kappa$ and center $\lambda(t)$. The friction coefficient is  $\gamma$, and $\xi(t)$ is a Gaussian random variable with zero mean and covariance $\ev{\xi(t)\xi(s)} = \delta(t-s)$. We denote time derivatives by overdots. Scaling all lengths by the equilibrium position standard deviation $\sigma \equiv \sqrt{\kT/\kappa}$ and times by the bead relaxation time $\tau_{\mrm{r}} = \gamma/\kappa$ gives the nondimensionalized Langevin equation
\begin{align}
    \dot{x}(t) = -\sr{x(t)-\lambda(t)} - \dg + \sqrt{2}\ \xi(t)\ ,
\label{eq:scaled_eom}
\end{align}
for scaled effective mass $\dg \equiv mg/(\kappa\sigma)$.

Integrating Eq.~\eqref{eq:scaled_eom} between measurements from $t$ to $t + \ts$ gives the discrete dynamics
\begin{align}
    \xkp = \xk\e^{-\ts} + \pr{1-\e^{-\ts}}\pr{\lk-\dg} + \sigma_{\ts}\xi_{k}\ ,
\label{eq:discrete}
\end{align}
for $\xk \equiv x(k\ts)$ and $\lk \equiv \lmd(k\ts)$. The variance $\sigma_{\ts}^{2} \equiv 1- \e^{-2\ts}$ of thermal force fluctuations depends on the sampling interval $\ts$, and $\xi_k$ is a Gaussian random variable with zero mean and covariance $\ev{\xi_{k}\xi_{n}}=\delta_{kn}$.

The effect of measurement noise is modeled by a measurement variable $y_{k}$ that is the sum of the bead's true position and additive white Gaussian noise,
\begin{align}
    \yk = \xk + \ssm \nu_k \ ,
\end{align}
with $\nu_k$ a Gaussian random variable with zero mean and covariance $\ev{\nu_k\nu_n} = \delta_{kn}$. We also assume that the thermal noise affecting the bead's position is independent of the measurement noise: $\ev{\xi_k\nu_n} = 0$, for all $k$ and $n$.

The trap position $\lk$ is updated at each time step using a ``ratcheting rule,''
\begin{align}
    \lkp = \lk + \Theta\pr{z_k-\lk}\sr{\alpha (z_k-\lk)},\ 
\label{eq:feedback_rule}
\end{align}
where $\Theta\pr{\cdot}$ denotes the Heaviside function, $\alpha$ the feedback gain, and $z_k = \{y_k,\hat{x}_{k+1}\}$ the estimate of the bead's position (using either the naive measurement $y_k$ or the Bayesian estimate $\hat{x}_{k+1}$). Because of delays, both the naive and Bayesian estimates of position use information from $y_k$ but not $y_{k+1}$; however, the Bayesian estimate also implicitly incorporates past information $\{y_{k-1}, y_{k-2},\ldots \}$ and uses the deterministic component of system dynamics to \textit{predict} $y_{k+1}$ (see Eq.~\eqref{eq:Bayesian_filter}, below).

\textit{Estimating the bead position}.---Since the actual bead position $x$ fluctuates on a scale $\sigma$ and the measurement noise fluctuates on a scale $\ssm$, it is convenient to define a signal-to-noise ratio $\snr \equiv \sigma/\sigma_{\mrm m}$~\footnote{SNR is often alternately defined as a ratio of signal-to-noise power, $\sigma^2/\sm$.}. In a first, \emph{naive} approach to designing feedback based on noisy measurements, the feedback rule, Eq.~\eqref{eq:feedback_rule}, directly uses the measurement $y_k$ to update the trap position $\lambda_{k+1}$. Notice that this method implicitly estimates the position $x_{k+1}$ by $y_k$. The naive method performs well at high SNR~($\gg 1$) but poorly at low SNR~($\ll 1$), where a unidirectional ratchet (implemented via the Heaviside function) often responds to noise rather than actual bead movements.  

In a second, \emph{filtering} approach, we improve the estimate of $x_{k+1}$ by using a Fokker-Planck equation to predict the position probability $p(x_{k+1})$ given $p(x_k)$, which is itself calculated from measurements up to $y_{k-1}$. One then updates (or ``corrects'') the prediction for time $k+1$ by incorporating the measurement $y_k$, using Bayes' rule~(see Appendix). For systems evolving according to linear dynamics and subject to Gaussian noise, $p(x_k)$ remains Gaussian for all $k$ (if $p(x_0)$ is initialized as Gaussian, see Appendix) and can be summarized by update equations for the mean $\hat{x}_k$ and variance. The Bayesian filter is then known as the \textit{predictive Kalman filter}, and one finds [compare Eq.~\eqref{eq:discrete}], 
\begin{align}
    \hat{x}_{k+1} = \underbrace{\hat{x}_{k}\e^{-\ts} + \pr{1-\e^{-\ts}}\pr{\lk-\dg}}_{\rm predict} + \underbrace{L\pr{\yk-\hat{x}_{k}}}_{\rm correct} \ ,
\label{eq:Bayesian_filter}
\end{align}
where the \textit{filter gain} $L$ ``corrects'' the naive prediction using the difference between the previous estimate and actual observation (see Appendix and Ref.~\cite[Ch.~8]{bechhoefer_book2021}). The gain $L$ is chosen to minimize the variance $\ev{(x_k-\hat{x}_k)^2}$ between the true position and its estimate, and the resulting value is a function of the SNR. This variance is always less than that of the measurement noise $\ssm^2$; it is optimal in that it incorporates all relevant past information contained in the (long) time series $\{y_k, y_{k-1}, \ldots \}$, and no other unbiased estimator---linear or not---has lower variance~\cite{kailath00}.

\textit{Engine thermodynamics}.---Given an estimate of the bead's position, we infer the thermodynamic quantities that characterize this engine's performance. The rate at which we extract gravitational energy (i.e., change in bead free energy) during the time interval $[t_k,t_{k+1})$ is~\cite{lucero2021maximal} 
\begin{align}
    \Delta F_{k+1} = \dg\pr{\lkp - \lk} \,,
\label{eq:grav_pow_est}
\end{align}
and the time-averaged rate of free-energy change (more informally, the \textit{output power}) is $\dot{F}=\tau^{-1}\sum_k \Delta F_k$, where $\tau = N\ts$ is the total duration of an $N$-step protocol. We estimate the incremental input work (of the trap on the bead) as
\begin{align}
    \Delta W_{k+1} = \frac{1}{2}\sr{\pr{\ykp - \lkp}^{2} - \pr{\ykp - \lk}^{2}} \,,
\label{eq:trap_pow_est} 
\end{align}
which estimates input work based on the noisy measurement $y_k$ and not on the true position $x_k$. The Appendix shows that this input-work estimator is unbiased---as a result of feedback delay.

\begin{figure}[tbp] 
    \centering
    \includegraphics[width=\linewidth]{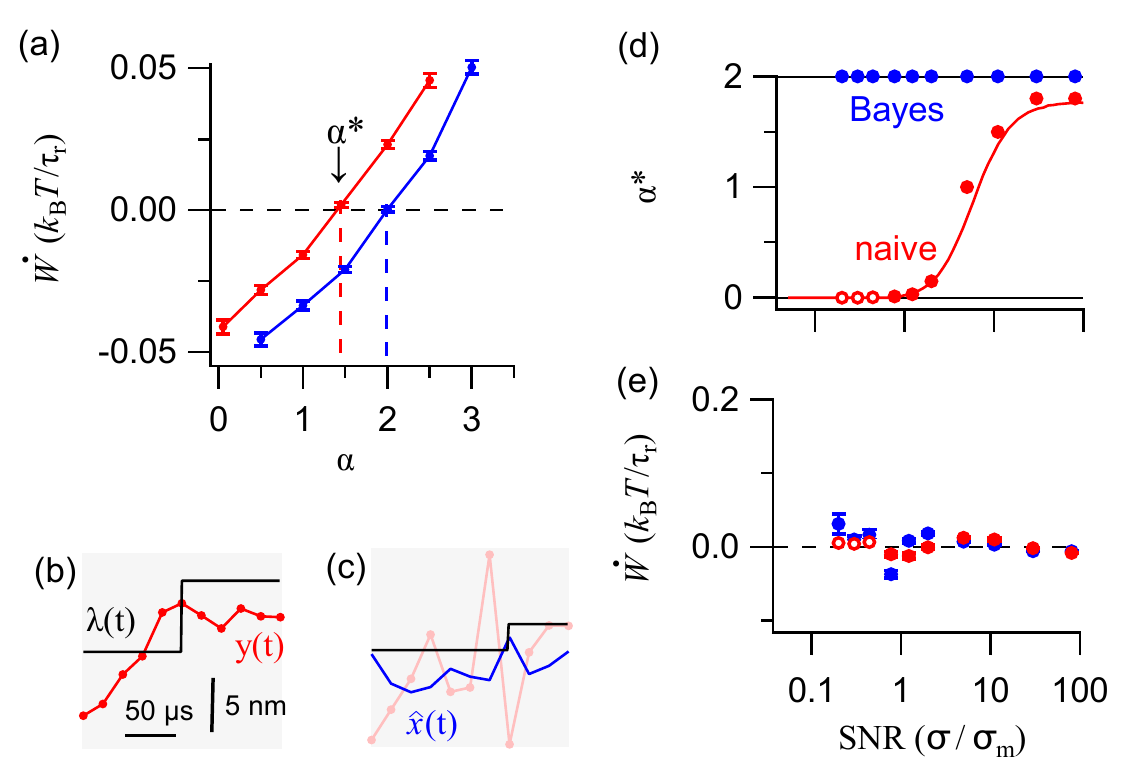} 
    \caption{Tuning the feedback gain $\alpha$ to set trap power $\dot{W}=0$. (a) Trap power for naive (red) and Bayesian (blue) information engines at fixed SNR = 10. (b) Measured bead ($y(t)$, red) and trap ($\lambda (t)$, black) trajectories for the naive information engine at SNR = 10. (c) Measured bead ($y(t)$, light red), filtered bead estimate ($\hat{x}(t)$, blue), and trap ($\lambda (t)$, black) trajectories for the Bayesian information engine at SNR = 2. (b) and (c) have equal scale bars and satisfy $\dot{W} = 0$. (d) Critical feedback gain $\astar$ and (e) corresponding input trap power, for naive (red) and Bayesian (blue) information engines. Hollow red markers denote SNRs for which $\astar > 0$ could not be found using the procedure outlined in (a). Solid red curve in (d) is from numerical simulation~(see Appendix). Experiments here and in other figures have sampling frequency $\trel/\ts = 41$, trap stiffness $\kappa = 42$~pN/\textmu m, scaled effective mass $\dg = 0.8$, diffusion constant $D = 0.12$~\textmu m$^2$/s, relaxation time $\trel = 0.8$ ms, and bead diameter 4~\textmu m. Markers denote experimental means, and error bars the standard errors of the mean.}
    \label{fig:PtrapVsalpha}
\end{figure}

\textit{Results}.---A ``pure'' information engine has zero input trap power: $\dot{W}~=~\tau^{-1}\sum_k \Delta W_k = 0$. 
For error-free measurements and no feedback delays, this criterion is satisfied at the critical feedback gain $\astar = 2$~(see Appendix and Ref.~\cite{saha2021maximizing}). 
Physically, this corresponds to translating the trap to a position opposite its minimum, so that the trap energy is unchanged. 
Feedback delay and measurement noise reduce $\astar$. 
For our experimental conditions with delay of one time step and SNR = 10, $\dot{W}=0$ is satisfied at the lower feedback gain $\astar \approx 1.5$ [Fig.~\ref{fig:PtrapVsalpha}(a-b)]. 

The Bayesian information engine applies feedback based on the filtered predictive estimate of the bead's position. As such, it accounts in its internal model for feedback delays and measurement noise. 

Figure~\ref{fig:PtrapVsalpha}(a) shows the input trap power, at fixed SNR~(=10), as a function of feedback gain $\alpha$.  Despite the delay and finite SNR, the input trap power is zero for the ``ideal'' feedback gain $\astar = 2$. Figure~\ref{fig:PtrapVsalpha}(c) illustrates the operation of the Bayesian filter with a trajectory of the Bayesian information engine at lower SNR ($=2$). In contrast to the naive information engine, the trap ratchets only when the estimated position (blue) crosses the trap center (black), and not necessarily when the noisy measurement (light red) crosses the trap center. 

Next, we investigate how $\astar$ depends on SNR, for the naive and Bayesian information engines. For the naive engine, $\astar$ decreases drastically at low SNR [Fig.~\ref{fig:PtrapVsalpha}(d)]. 
For $\snr \lesssim \snrc = 0.7 \pm 0.1$, where $\snrc$ denotes the critical vale of the SNR, a non-zero $\astar$ could not be found experimentally using the procedure outlined in Fig.~\ref{fig:PtrapVsalpha}(a). 
The Appendix shows that the vanishing of $\astar$ corresponds to a kind of phase transition between a regime where one can set $\dot{F} > 0$ while maintaining $\dot{W}=0$ and a regime where one cannot.

By contrast, the critical feedback gain $\astar$ remains near 2 for the Bayesian engine [Fig.~\ref{fig:PtrapVsalpha}(d)]. The corresponding measured input trap powers for both the naive and Bayesian information engines are close to zero relative to the maximum output power ($\dot{F}_\mrm{max}\approx 0.27\, \kT/\trel$) of the engine, at all SNR [Fig.~\ref{fig:PtrapVsalpha}(e)]. 

Taking advantage of predictions in our estimation algorithm thus simplifies the experiments, as it eliminates the need to empirically tune the feedback gain, ensuring that the zero-work condition is always satisfied at $\alpha = 2$. 
Above, we saw that it also simplifies the work calculations needed to realize a pure information engine, as the value calculated directly from the noisy measurement is an unbiased estimator of the true work.

\begin{figure}[htbp] 
    \centering
    \includegraphics[width=1.0\linewidth]{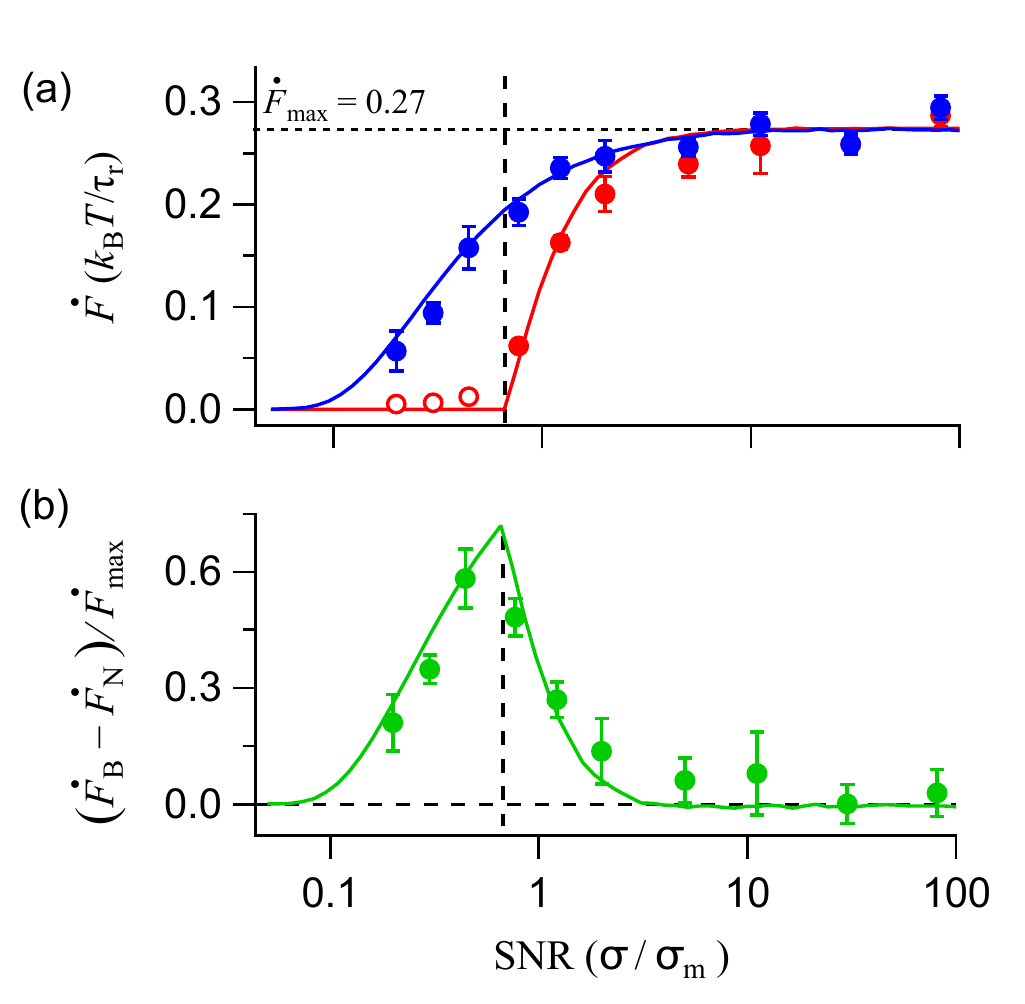} 
    \caption{Performance of the information engines. a) Output power of naive (red) and Bayesian (blue) information engines as a function of SNR. Hollow red markers denote output power at $\astar = 0$. b) Difference of output work extraction rates for the Bayesian (B) and naive (N) engines scaled by the maximum rate ($\dot{F}_\mrm{max} = 0.27$). The difference peaks at SNR=$\mrm{SNR}_\mrm{c}$~$\approx 0.7$ (vertical dashed lines). Markers denote experimental means, solid curves the numerical simulations~(see Appendix). Error bars denote a) standard error of the mean and b) propagated standard error of the mean from a).}
    \label{fig:OutputPower}
\end{figure}

Finally, Fig.~\ref{fig:OutputPower}(a) compares the performance of the naive and Bayesian information engines, as quantified by the rate of stored gravitational power $\dot{F}$ while keeping $\dot{W} = 0$. Both output powers $\dot{F}$ increase monotonically with SNR and saturate at the same power level at high SNR ($> 10$).  

Although the performance of Bayesian and naive information engines is similar at low and high SNR, there is a striking contrast at intermediate $\snr \lesssim 1$. Indeed, the difference of output powers (Bayesian$-$naive), normalized by $\dot{F}_\mrm{max}$, significantly exceeds zero for $0.1~\leq~\mrm{SNR}~\leq 2$ and reaches a maximum at $\mrm{SNR} \approx \snrc$ (Fig.~\ref{fig:OutputPower}b).

At high SNR, the Bayesian filter ``trusts the observation'' and returns an estimate close to the instantaneous measurement, corrected for the expected bias due to the time delay. Since this bias is small for frequent measurements, both engines have similar performance and extract all the favorable thermal fluctuations, saturating at the maximum output power $\dot{F}_\mrm{max} \approx 0.27$. At low SNR, the measurements are so noisy that they exceed the scale of the trap.  The Bayesian information engine then extracts negligible power~(see Appendix), while the naive engine extracts zero power. Therefore, at SNR $\gg 1$ and $\ll 1$, the difference of output powers ($\dot{F}_\mrm{B} - \dot{F}_\mrm{N}$) tends to zero. But at intermediate SNR, the effective noise averaging in the Bayesian (Kalman) filter produces more accurate estimates, leading to better feedback decisions and thus improved engine performance. 

To understand why the naive information engine shows a phase transition at a critical signal-to-noise ratio, we numerically solve a self-consistent equation for the SNR at which trap power vanishes. Enforcing the condition that $\astar = 0$ is the unique solution (see Appendix), we find $\snrc \approx 0.64$, consistent with both numerical simulations and experiments. 
 
We also find that the phase transition arises from the \emph{biased} estimate of the bead's position from the noisy measurements. This bias has two origins: the delay due to feedback latency and the failure of the naive measurement to account for the fact that fluctuations above threshold are rare while noise fluctuations of either sign are equally likely. Because fluctuations up to the threshold are rare, the bead is usually below the observed value whenever an apparent
threshold crossing is observed.

By contrast, a phase transition does \textit{not} occur for the Bayesian information engine. The Bayesian filter gives an \textit{unbiased} prediction of the bead's position, accounting for both feedback delay and the ``prior'' associated with observations near the threshold. As a result, the bead is equally likely to be on either side of the predicted position, allowing one to tune for zero trap power and extract at least some power at \textit{any} SNR value (see Appendix).

\textit{Conclusion}.---Information engines that decide whether to ratchet using single noisy measurements have a phase transition at a critical signal-to-noise ratio $\snrc$ and cannot function for $\snr < \snrc$. By contrast, if its feedback uses a Bayesian estimate of bead position that incorporates prior measurements, an information engine can operate at \textit{all} values of SNR. The maximum performance benefit over the naive engine occurs at the critical value $\snrc$. 

The ability to increase the performance of an information engine at low SNR is important for experimental investigations of motor mechanisms that use fluorescent probes~\cite{veigel11}. In such applications, lower light intensities for monitoring fluorescent probes reduce photobleaching and allow longer measurements of motor behavior.

In addition, using a filtering algorithm to reduce the required accuracy of information while maintaining a given performance may decrease the thermodynamic costs of processing position measurements. Generally, a lower measurement accuracy reduces the minimum thermodynamic costs of running the controller~\cite{Horowitz2014_Second-law_like,paneru2020efficiency}. However, keeping a memory of past observations should increase those costs. Further work is needed to understand the efficiency of a feedback strategy that incorporates a memory of past measurements with one based purely on the most recent measurement. Such studies could evaluate the potential performance trade-offs encountered when varying the measurement accuracy.
\begin{acknowledgments}
This research was supported by grant FQXi-IAF19-02 from the Foundational Questions Institute Fund, a donor-advised fund of the Silicon Valley Community Foundation. Additional support was from Natural Sciences and Engineering Research Council of Canada (NSERC) Discovery Grants (D.A.S.\ and J.B.) and a Tier-II Canada Research Chair (D.A.S.), an NSERC Undergraduate Summer Research Award, a BC Graduate Scholarship, and an NSERC Canadian Graduate Scholarship -- Masters (J.N.E.L.).
\end{acknowledgments}
\section*{Appendix}
\appendix
\beginsupplement
\section{Experimental apparatus}
\label{sec:expt_apparatus}

The experimental apparatus is designed to optically trap a bead in water. The optical trap is built by focusing a green laser using a water-immersion objective (Olympus, 60x 1.2 NA). The position of the trap is controlled using an acousto-optic deflector (AOD, AA opto-electronics). The position of the trapped bead is measured by a red detection laser, propagating anti-parallel to the trapping laser, that is loosely focused in the trapping plane using an air objective. The intensity of the detection bead is controlled using an acousto-optic modulator (AOM, AA opto-electronics) to vary the measurement noise. More details about the experimental apparatus can be found in \cite{saha2021maximizing,kumar2018nanoscale}. 

\begin{figure}[ht]
    \centering
    \includegraphics[width=1\linewidth]{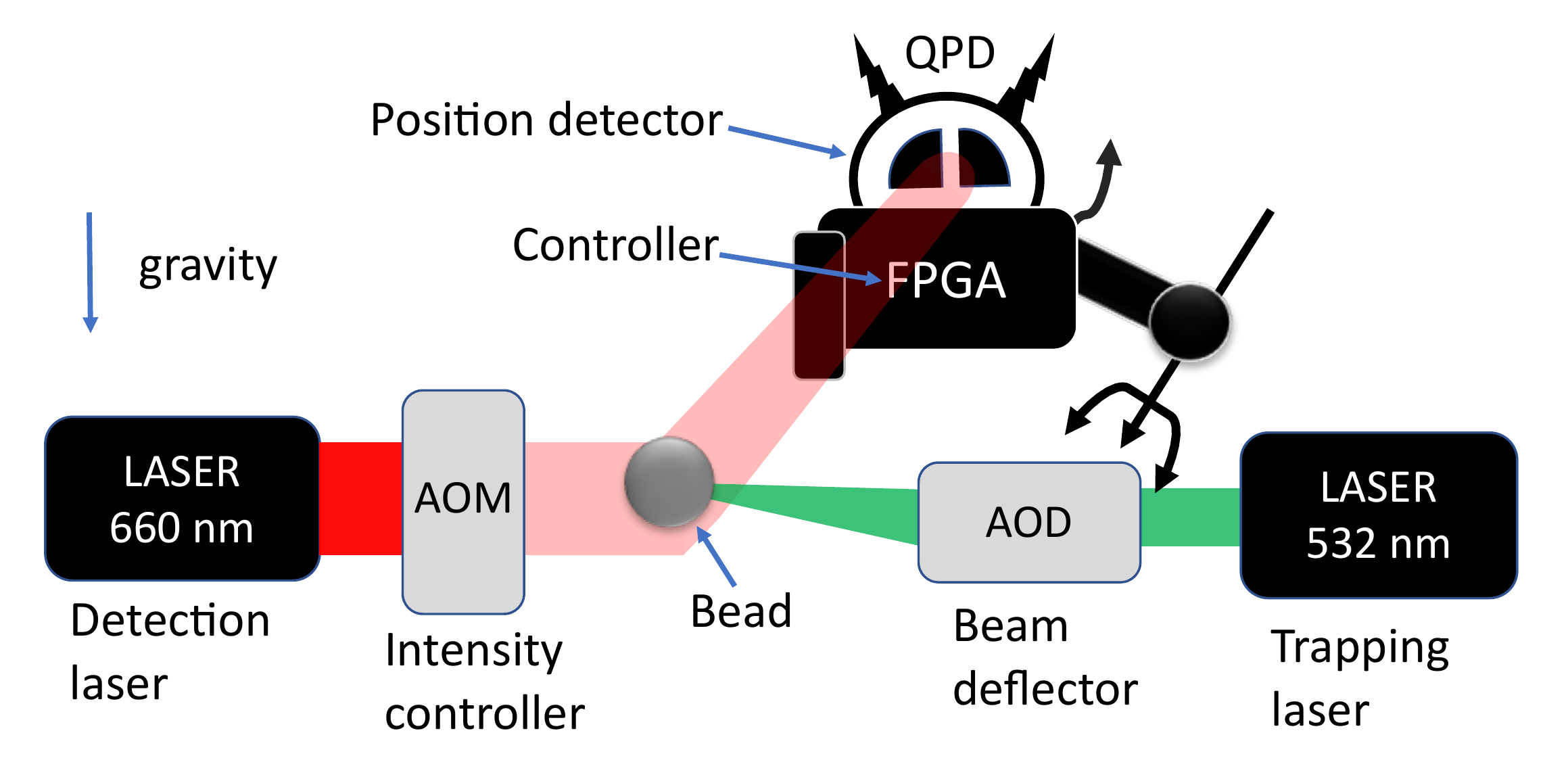}
    \caption{Schematic diagram of the experimental apparatus.}
    \label{fig:Expt_apparatus}
\end{figure}

\section{Experimental parameter estimation}
The trap stiffness, diffusion constant, and measurement noise are measured by fitting the power spectrum of the bead's position to a discrete aliased Lorentzian~\cite{berg2004power}. The discrete equation of motion of the bead can be re-written as (cf.\ Eq.~(3) in the main text)
\begin{align}
    x_{k+1} = c\, x_k + \Delta x\, \xi_k \,,
\end{align}
where $x$ is measured with respect to $\lambda - \dg$. The parameters are
\begin{align}
    c \equiv \exp(-\ts/\trel) = \exp(-2\pi\fc \ts) \,,
\end{align}
and
\begin{align}
    \Delta x \equiv \left[\frac{(1-c^2)D}{2\pi\fc} \right]^{1/2} \,,
\end{align}
where $D$ is the diffusion constant of the bead, $\fc$ the corner frequency, and $\ts$ the sampling time. Applying the discrete Fourier transform to the discrete equation of motion of the bead and calculating the expected value of the power spectrum $P_f$, the discrete aliased Lorentzian is
\begin{align}
    P_f^x = \frac{2(\Delta x)^2 \ts}{1+c^2-2c \, \cos 2\pi f \ts} \,,
\end{align}
where $f$ indexes the discrete non-negative frequencies. For noisy measurements 
\begin{align}
y = x+\nu \,,
\label{eq:noisy_meas}
\end{align} where the measurement noise is Gaussian white noise, the power spectrum of $y$ is 
\begin{align}
    P_f^y = P_f^x+P^\nu \,,
    \label{eq:discrete_lorr_noise}
\end{align}
where $P^\nu$ is the frequency-independent noise floor, and $\sigma_m^2 = P^\nu/2\ts$ the variance of the measurement noise. Equation~\eqref{eq:discrete_lorr_noise} uses the property that the measurement noise is independent of both the bead's position $x$ and thermal noise~$\xi$. 

\begin{figure}[ht]
    \centering
    \includegraphics[width=0.65\linewidth]{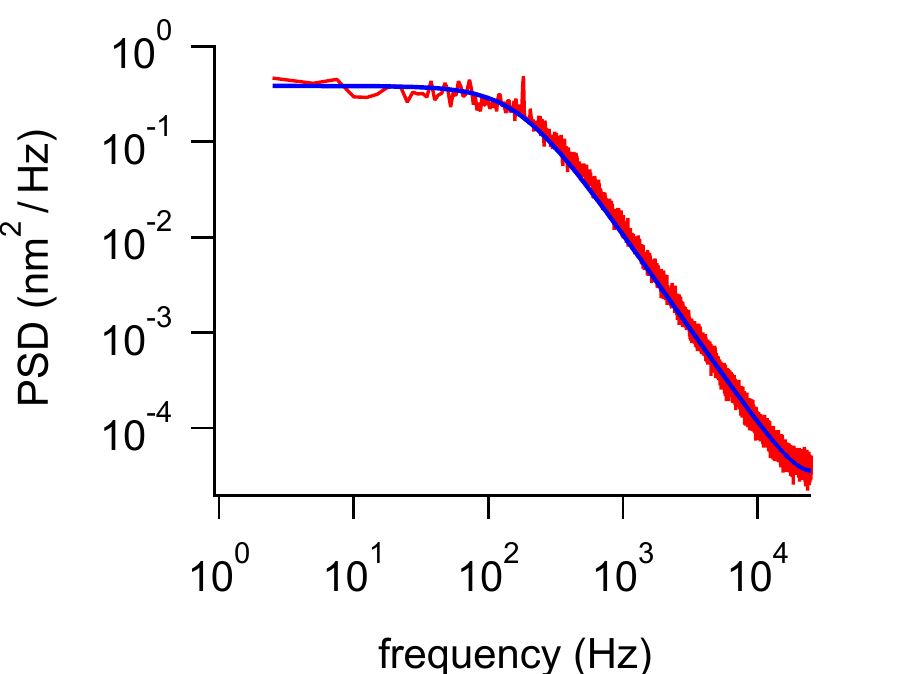}
    \caption{
    Power spectral density $P_{f}^{y}$ of the measured bead position $y$ in a static trap. Red curve: experiment; blue curve: fit to \eqref{eq:discrete_lorr_noise}. The measured fit parameters are $f_\mrm{c} = 167.8\pm1.6$ Hz, $D = 0.114\pm 0.001$ \text m$^2$/s, and $\sigma_\mrm{m} = 0.42\pm0.06$ nm.}
    \label{fig:PSD}
\end{figure}

Figure~\ref{fig:PSD} shows the power spectral density of a 4-\textmu m bead in a static trap and the fit to \eqref{eq:discrete_lorr_noise}. The trap stiffness is estimated using $\kappa = 2\pi f_\mrm{c}k_\mrm{B}T / D = 40$ pN/\textmu m.

\section{Performance of Bayesian-filter estimator} 

When the measurement noise is additive Gaussian white noise and the system dynamics are linear, as is closely approximated in this experimental setup, an exactly solvable Bayesian filter can be designed to minimize the least-squares error between the filter's state prediction and the true bead position. To derive the optimal filter for the information engine, we note that the bead's motion is a first-order Markov process: its next position $\xkp$ is conditionally independent of its history given its current position $\xk$ and the trap position $\lambda_k$, $p\pr{\xkp|\xk,\xkm,\dots,x_0,\lambda_k} = p\pr{\xkp|\xk,\lambda_k}$. Here, the probability $p\pr{\xkp|\xk,\lambda_k}$ is obtained from Eq.~(3),
\begin{align}
    p &\pr{ \xkp |\xk,\lk } \sim \label{eq:OU_dynamics} \\
    &\mcal{N} \pr{ \xkp;\xk \e^{-\ts} + \left( 1-\e^{-\ts} \right) \left( \lambda_k -\dg \right), \, 1-\e^{-2\ts} } \ , \nonumber
\end{align}
where $\mcal{N}(x;\mu, \sigma^{2})$ denotes a normal distribution over $x$ with mean $\mu$ and variance $\sigma^{2}$. We also assume that the measurements of the bead's position are \emph{memoryless}: the measurement at the current time depends only on the current state, $p\pr{\yk|\xk, \xkm, \dots, x_0} = p\pr{\yk|\xk}$. Measurements with additive Gaussian white noise satisfy a distribution of the form
\begin{align}
    p\pr{\yk|\xk} \sim \mcal{N}\pr{\yk;\xk, \, \sigma_{\mrm m}^{2}}\ . \label{eq:meas_dist}
\end{align}
The standard deviation of the noise distribution is the reciprocal of the signal-to-noise ratio, $\ssm = 1/\snr$, and as such can be interpreted as a ``noise-to-signal'' ratio. Using Bayes' rule, the update distribution is
\begin{align}
    p\pr{\xk|\Yk} = \frac{p\pr{\yk|\xk}\, p\pr{\xk|\Ykm}}{p\pr{\yk|\Ykm}}\ ,
\end{align}
where $\Yk = \{\yk,\ykm,\dots,y_{1}\}$ is the set of measurements made up to and including time $t_{k}$, and $p\pr{\yk|\Ykm}$ is a normalizing distribution. From this updated distribution and the dynamics~\eqref{eq:OU_dynamics}, the predicted distribution at the next discrete time is
\begin{align}
    p\pr{\xkp|\Yk} = \int\dd{\xk}\, p\pr{\xkp|\xk}\, p\pr{\xk|\Yk}\ .
\end{align}
The measurement update and prediction steps for the information engine can be formulated as a sequence of transformations of univariate Gaussian probability densities. Since all densities are Gaussian, we need propagate only their means and variances:
\begin{widetext}
\begin{subequations}\label{eq:KF_equations}
    \begin{align}
        \left.
        \begin{aligned}
            \hat{P}_{k}^{y} &= \hat{P}_{k} + \sm&&\quad\textrm{obs. var.}\\
            L_{k+1} &= \e^{-\ts}\frac{\hat{P}_{k}}{\hat{P}_{k+1}^{y}}&&\quad\textrm{Kalman gain}\\
            \hat{P}_{k}^{+} &= - \pr{L_{k+1}}^{2}\hat{P}_{k+1}^{y}&&\quad\textrm{state var.}\\
            \hat{x}_{k}^{+} &= L_{k+1}\pr{y_{k}-\hat{x}_{k}}&&\quad\textrm{state mean}\\
        \end{aligned}
        \right\}
        \textrm{update}\label{eq:KF_update}\\
        \nonumber\\
        \left.
        \begin{aligned}
            \hat{P}_{k+1} &= \hat{P}_{k}^{+} + \e^{-2t_{\mrm s}}\hat{P}_{k} + \pr{1-\e^{-2t_{\mrm s}}}&&\quad\textrm{state var.}\\
            \hat{x}_{k+1} &= \hat{x}_{k}^{+} + \e^{-\ts}\hat{x}_{k} + \pr{1-\e^{-\ts}}\pr{\lk-\dg}&&\quad\textrm{state mean}\\
        \end{aligned}
        \right\}
        \textrm{predict}\label{eq:KF_predict}
    \end{align}
\end{subequations}
\end{widetext}

Intuitively, the filter generates a prediction mean and variance~\eqref{eq:KF_predict}, $\xhkp$ and $\hat{P}_{k+1}$. The stochastic bead dynamics increase the uncertainty of the bead's position. This position uncertainty is reduced by the information obtained from the previous measurement $\yk$. The reduction in prediction uncertainty and corresponding change to the prediction mean is achieved by feedback based on the discrepancy between the actual measurement and the prediction, $y_{k}-\xhk$. The \textit{Kalman gain} $L$ multiplying this discrepancy accounts for the unreliability of the measurements, given the scale of measurement noise. The information engine uses this filter's prediction mean as the estimate of the bead's position in the feedback decision.

\section{Steady-state Bayesian-filter gain}

In experiments, we assume that the experimental parameters do not change with time. As a result, the means and variances in the time-dependent Eqs.~\eqref{eq:KF_equations} quickly approach steady-state values. To obtain the steady-state Bayesian-filter gain, we first solve the associated Riccati equation~\cite[Ch. 8]{bechhoefer_book2021} for the predictive variance,
\begin{align}
    \hat{P}^{\mrm{SS}} &= \e^{-2\ts}\hat{P}^{\mrm{SS}} + \pr{1-\e^{-2\ts}} - \e^{-2\ts}\frac{\pr{\hat{P}^{\mrm{SS}}}^{2}}{\hat{P}^{\mrm{SS}} + \sm}\ , \label{eq:steady_filter_var}
\end{align}
where superscript ``SS'' without time subscript indicates a steady-state quantity. The relevant solution to this quadratic equation is
\begin{align} 
    \hat{P}^{\mrm{SS}} = \frac{1}{2}&\pr{1-\sm}\pr{1-\e^{-2\ts}}\label{eq:steady_filter_var_sol} \\
    &\times \Bigg[1 + \sqrt{1 + \frac{4\sm}{\pr{1-\sm}^{2}\pr{1-\e^{-2\ts}}}}\,\, \Bigg] \,. \nonumber
\end{align}
We discard the other solution, as it is negative and variances must be positive. The resulting steady-state Bayesian (Kalman) filter gain is
\begin{align}
    L^{\mrm{SS}} = \frac{\hat{P}^{\mrm{SS}}}{\hat{P}^{\mrm{SS}}+\sigma_m^2}\e^{-\ts}. 
\end{align}

Finally, the Kalman-filtered position estimate is
\begin{align}
    \hat{x}_{k+1} &= L^{\mrm{SS}}\pr{y_{k}-\hat{x}_{k}} + \e^{-\ts}\hat{x}_{k} + \pr{1-\e^{-\ts}}\pr{\lk-\dg}.
    \label{eq:KalmanEstimate}
\end{align}

\section{Experimental implementation of the Bayesian filter}

\begin{figure}[ht]
    \centering
    \includegraphics[width=0.70\linewidth]{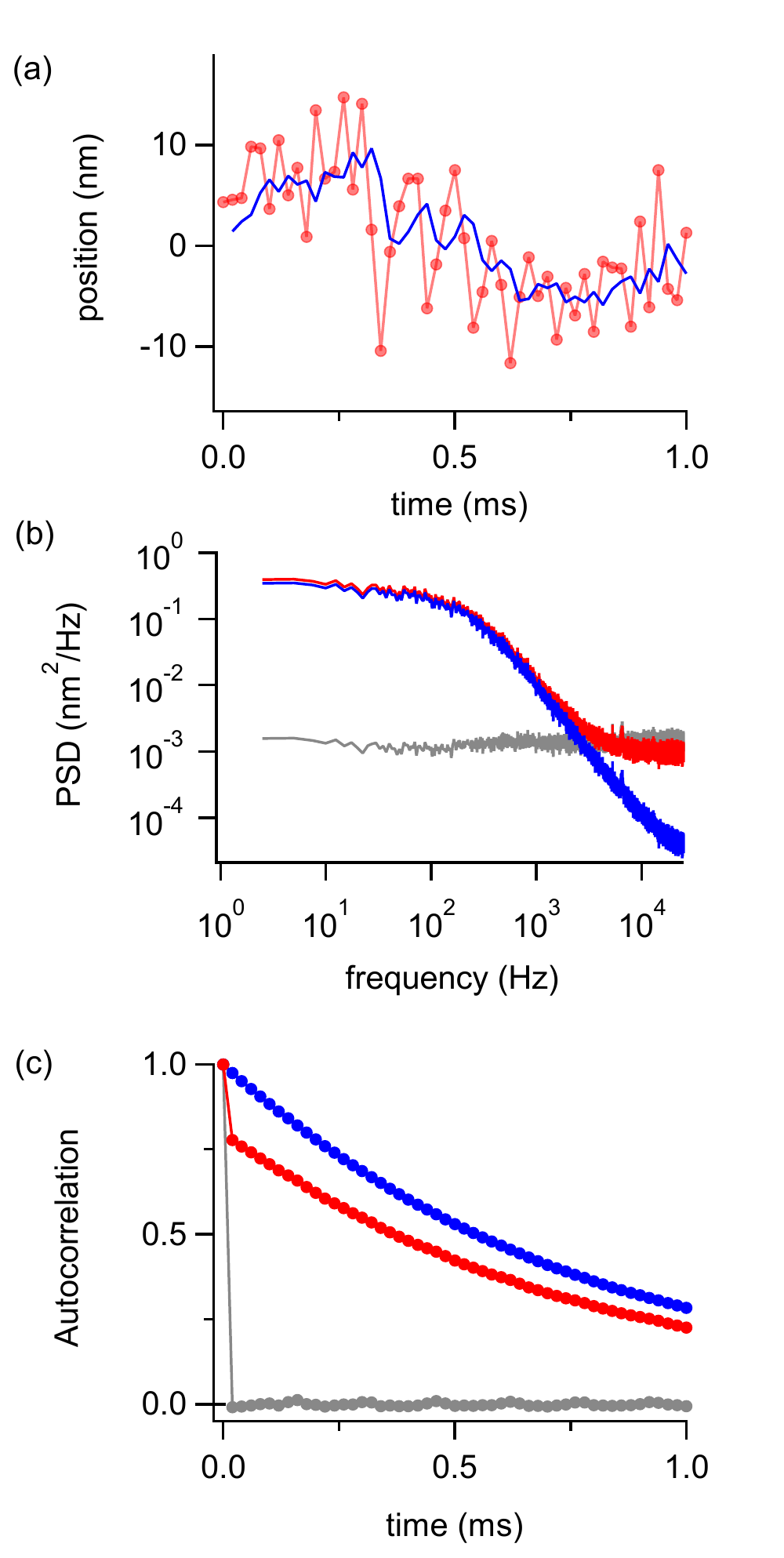}
    \caption{
    Experimental performance of the Bayesian-filtered estimate. (a) Noisy experimental measurement ($y$, red) and the corresponding Bayesian estimate using Eq.~\eqref{eq:KalmanEstimate} ($\hat{x}$, blue) of the bead's position. (b) Power spectral density and (c) autocorrelation of the noisy measurement (red), Bayesian estimate (blue), and innovations $\hat{x}-y$ (gray). The experiments were performed for SNR $\approx$ 2, equilibrium standard deviation $\sigma = 10.1$~nm, and measurement noise $\ssm = 4.7$~nm.}
    \label{fig:Autocorrelation_PSD}
\end{figure}

Experiments on the Bayesian information engine were performed using the Kalman-filtered position estimate obtained from Eq.~\eqref{eq:KalmanEstimate}. Figure~\ref{fig:Autocorrelation_PSD} shows the performance of the experimental implementation of the Kalman filter for a bead in a static trap. The experiment was performed at SNR $\approx$ 2, where the experimental parameters were estimated by fitting the power spectral density using Eq.~\eqref{eq:discrete_lorr_noise}. Figure~\ref{fig:Autocorrelation_PSD}(a) shows the trajectory of 
the estimated bead position (blue), inferred using Eq.~\eqref{eq:KalmanEstimate} from the noisy measurements (red).

Figure~\ref{fig:Autocorrelation_PSD}(b) shows the power spectral densities of the noisy measurements (red) and Bayesian estimates (blue). The power spectral density of the noisy measurements saturates at a finite noise floor (PSD$~=~10^{-3}$~nm$^2$/Hz). The filter-estimated position (blue) has a noise floor less then a tenth that of the individual noisy measurements (red). The power spectral density of the innovation $\hat{x}_{n} - y_{n}$ is flat, showing that the innovations follow a white-noise distribution, and hence all the position correlations arising from the bead's relaxation in the harmonic trap are extracted by the estimator.

Figure~\ref{fig:Autocorrelation_PSD}(c) shows the normalized autocorrelation of the noisy measurements (red), Bayesian estimates (blue), and innovations (gray). From Eq.~\eqref{eq:noisy_meas}, the autocorrelation of the noisy measurements is 
\begin{subequations}
\begin{align}
    \ev{y_ny_m} &= \ev{(x_n+\nu_n)(x_m+\nu_m)} \\
    &= \sigma_\mrm{m}^2\delta_{nm} +\e^{(m-n)\ts},
    \label{eq:noisy_autocorrelation}
\end{align}
\end{subequations}
where we have again used the property that the measurement noise is independent of the true bead position. 

The noisy measurements (red curve) show a delta-function peak at $t=0$ on top of a broader exponential decay, reflecting two processes:  First, there is the contribution of measurement noise, which is delta correlated; second, there is the contribution of the dynamics, which has an exponential decay of correlations. 

By contrast, the Bayesian estimate (blue curve) has only an exponential decay due to the dynamics---as would the autocorrelation function of noise-free dynamics. In other words, the Bayesian estimator distills the dynamical information in the correlated signal. The innovations (gray curve) are almost entirely delta correlated.  Indeed, the degree to which the innovations approximate a delta function serves as a measure of the quality of the Bayesian filter. For example, deviations between a parameter value in the physical system and the corresponding model used for the Bayesian filter produce dynamical correlations in the innovations autocorrelation function~\cite{bechhoefer_book2021}. Indeed, this mismatch has been used as a way to empirically tune parameters for Bayesian (Kalman) filters (``innovation whitening")~\cite{wang11}.

\section{Numerical-simulation methods}

The numerical simulations generate multiple long trajectories of the bead using the discrete-time dynamics,
\begin{align}
    \xkp = \xk\e^{-\ts} + \pr{1-\e^{-\ts}}\pr{\lk-\dg} + \sigma_{\ts}\xi_{k}\ ,
\end{align}
where $\xk \equiv x(k\ts)$ and $\lk \equiv \lmd(k\ts)$. The corresponding noisy measurement is $\yk = \xk + \ssm \nu_k$, where $\nu_k$ is a Gaussian random variable with zero mean and covariance $\ev{\nu_k\nu_n} = \delta_{kn}$. For the naive information engine, the trap position is updated using the feedback rule 
\begin{align}
    \lkp = \lk + \Theta\pr{y_k-\lk}\, \alpha (y_k-\lk) .
\end{align}
For the Bayesian information engine, the trap position is updated using the feedback rule
\begin{align}
    \lkp = \lk + \Theta\pr{\hat{x}_{k+1}-\lk}\, \alpha (\hat{x}_{k+1}-\lk)\ ,
\end{align}
using the Bayesian estimate~\eqref{eq:KalmanEstimate} of the bead's position. For each engine type, the free-energy change is estimated using Eq.~(7) and the work using the unbiased estimator Eq.~(8) in the main text.

\section{Trap-work estimates are unbiased} 
Here, we show that our empirical estimator for the input trap work 
\begin{align}
    \wt_{k+1} = \frac{1}{2}\sr{\pr{\ykp-\lkp}^{2} - \pr{\ykp-\lk}^{2}}
\label{eq:naive_emp_trap_pow}
\end{align}
is an unbiased estimator when the feedback rule follows 
\begin{align}
    \lkp = \lk + \Theta\pr{z_k-\lk}
    \, \alpha (z_k-\lk)
    \ .
\end{align}
Note that this feedback rule updates the value of the control parameter $\lk\to \lkp$ at time $t_{k+1}$ according to the estimate $z_k$ of the bead's position at the \emph{previous} time $t_{k}$, not the measurement that is made at the current time $t_{k+1}$. In other words, there is a delay of one cycle between when a measurement is taken and when the corresponding feedback is applied.

For convenience, we introduce a new notation where the superscript on the thermodynamic quantities (e.g., $\wt_{k}^{x}$) denotes the type of measurement used in the estimation of the thermodynamic quantity. In this notation, we write \eqref{eq:naive_emp_trap_pow} as
\begin{align}
    \wt_{k+1}^{y} = \frac{1}{2}\sr{\pr{\ykp-\lkp}^{2} - \pr{\ykp-\lk}^{2}}\ ,
\end{align}
and the true estimate of the work from stochastic thermodynamics~\cite{sekimoto2010} given by the actual potential-energy change when updating $\lambda_k$ to $\lambda_{k+1}$,
\begin{align}
    \wt_{k+1}^{x} = \frac{1}{2}\sr{\pr{\xkp-\lkp}^{2} - \pr{\xkp-\lk}^{2}}\ .\label{eq:true_energy_est}
\end{align}

\begin{figure}[!htbp]
    \centering 
    \includegraphics[clip, width=0.75\linewidth]{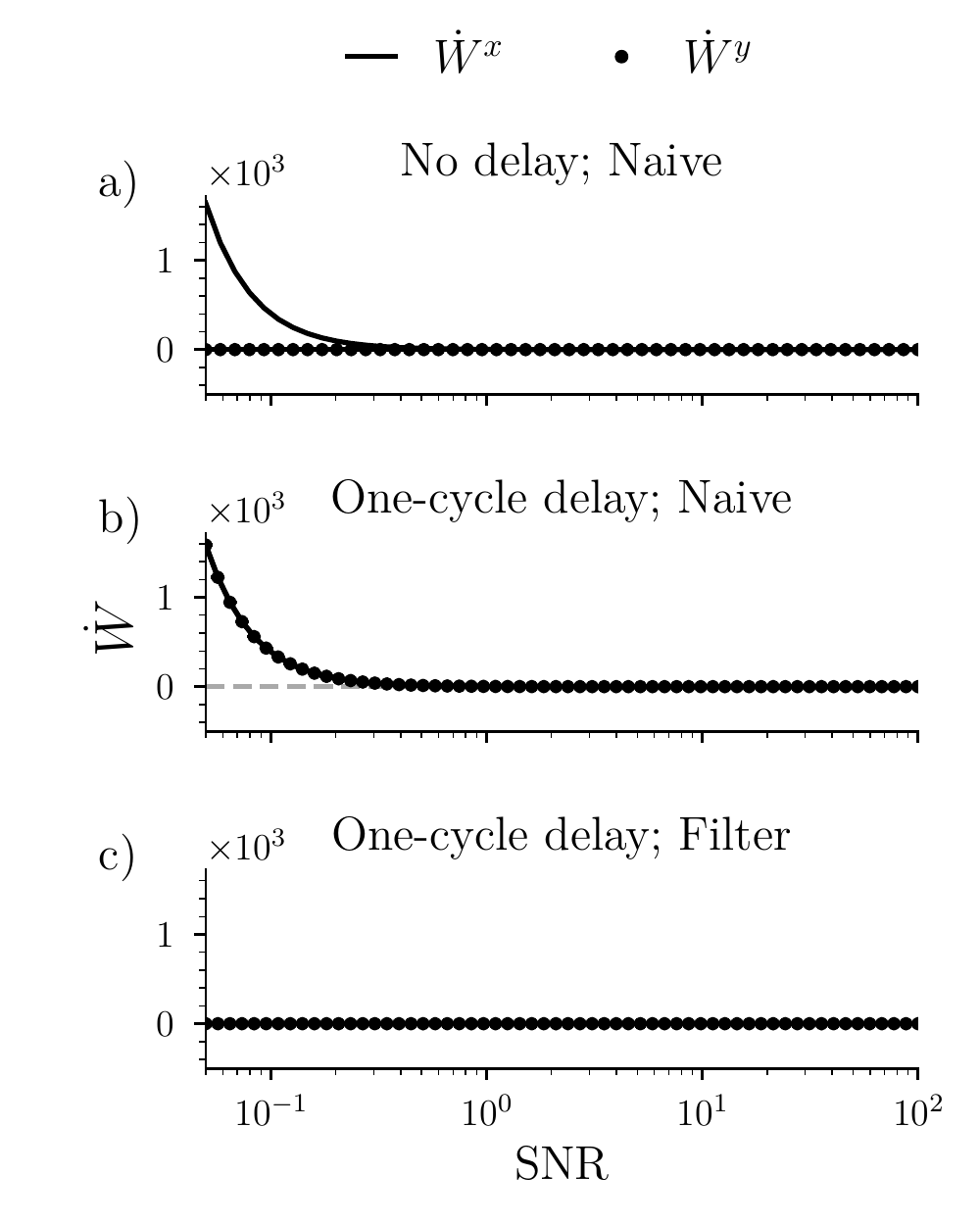}
    \caption{Numerical simulations of bias in the trap-power estimates from noisy measurements, as a function of SNR. a) Estimated trap power in naive information engine with no feedback delay (latency) and applying feedback based on the most recent noisy measurement: $\lkp = \lk + \Theta\pr{y_{k+1}-\lk}\sr{\alpha (y_{k+1}-\lk)}$. b), c) Estimated trap power when feedback is applied after a delay of one time step, using the feedback rule~(5) in the main text, based on 
    (b) biased naive $z_k = y_k$ or (c) unbiased filtered $z_k = \hat{x}_{k+1}$ position estimates. Solid curve: estimate using true positions $x$. Markers: estimate using measurements $y_{k}$. The numerical simulations were performed for sampling frequency $\fs = 40$, feedback gain $\alpha = 2$, and scaled mass $\dg=0.84$.}
    \label{fig:naive_with_wo_delay}
\end{figure}

Now we compare the empirical estimates to the true values of the thermodynamic quantities. Expanding the definition of the empirical input work~\eqref{eq:naive_emp_trap_pow} using $y_k = x_k + \nu_k$, where $\nu_k$ is a zero-mean Gaussian random variable with variance $\sigma^2_\mathrm{m}$, gives
\begin{subequations} 
    \begin{align}
        \wt_{k+1}^{y} &= \frac{1}{2}\sr{\pr{\ykp-\lkp}^{2} - \pr{\ykp-\lk}^{2}}\\
        &= \frac{1}{2}\left[\pr{\xkp-\lkp + \nu_{k+1}}^{2} \right.\\
        &\hspace{3cm}\left.-\pr{\xkp-\lk+\nu_{k+1}}^{2}\right]\nonumber\\
        &\hspace{-0.5cm}= \frac{1}{2}\sr{\pr{\xkp-\lkp}^{2} - \pr{\xkp-\lk}^{2}} \\
        &\hspace{4cm}+ \pr{\lkp-\lk}\nu_{k+1}\nonumber\\
        &= \pr{\wt_{k+1}}^{x} + \pr{\lk-\lkp}\nu_{k+1}\ .
    \end{align}
\end{subequations}
Averaging over the noise at time $t_{k+1}$, when the trap position has a nonnegative update $\lkp-\lk > 0$ (when $\lkp=\lk$, the work is trivially $\wt = 0$ regardless of which position estimate is used), we find that 
\begin{subequations}
    \begin{align}
        \ev{\wt_{k+1}^{y}}_{\nu} &= \ev{\wt_{k+1}^{x}}_{\nu} + \ev{\pr{\lk-\lkp}\nu_{k+1}}\\
        &\hspace{-1.75cm}= \ev{\wt_{k+1}^{x}}_{\nu} + \ev{\sr{\lk-(\lk+\alpha(z_k-\lk))}\nu_{k+1}}\\
        &\hspace{-1.75cm}= \ev{\wt_{k+1}^{x}}_{\nu} - \alpha\ev{z_k\ \nu_{k+1}} +\alpha\ev{\lk\ \nu_{k+1}}\\
        &\hspace{-1.75cm}= \ev{\wt_{k+1}^{x}}_{\nu}\ ,
    \end{align}
\end{subequations}
where, in the last step, we have used the property that both the position estimate $z_{k}$ (which denotes the noisy measurement $y_k$ in the naive case and the predictive estimate $\xhkp$ in the Bayesian case), and the trap position $\lk$ at time step $k$ are independent of the \emph{future} measurement noise $\nu_{k+1}$. Without the delay, the variables are correlated and the measurements are biased (Fig.~\ref{fig:naive_with_wo_delay}a). This calculation suggests that feedback delay actually eliminates the bias in the empirical estimate of the trap work. Using numerical simulations, we compute the empirical power $\dot{W}^{y}$ and find that it agrees well with the true values $\dot{W}^{x}$ for both the naive (Fig.~\ref{fig:naive_with_wo_delay}b) and Bayesian (Fig.~\ref{fig:naive_with_wo_delay}c) information engines.

We also note that the trap work could alternatively be estimated using the Bayesian filtered position; however, the estimator in Eq.~\eqref{eq:naive_emp_trap_pow} is \textit{robustly unbiased}: It is unbiased even if the dynamical model used in the Kalman filter uses parameters that differ from those describing the physical system. The calculation of the trap position $\lambda_k$ is based on past information and---whatever the algorithm used to compute its value---will always be independent of the current measurement.

\section{Experiment and simulation of the critical feedback gain $\astar$}

\begin{figure}[thbp]
    \centering
    \includegraphics[clip, width=0.65\linewidth]{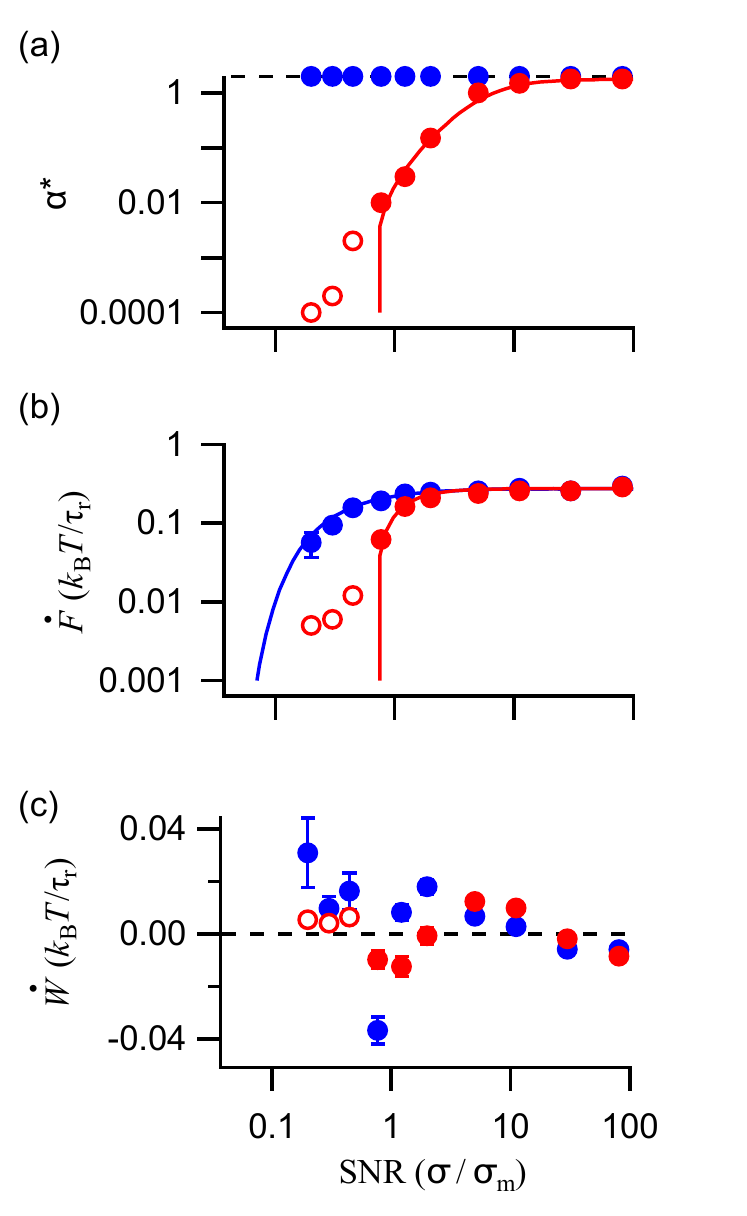}
    \caption{Experiments (markers) and simulations (curves) from Fig.~2 in the main text, presented on a logarithmic ordinate. (a) The critical feedback gain $\astar$, and the corresponding (b) output power $\dot{F}$ and (c) trap power $\dot{W}$ of the naive (red) and Bayesian (blue) information engines, each as a function of SNR. 
    Hollow markers denote experiments at $ \snr < \snrc$. Error bars denote the standard error of the mean.
    }
    \label{fig:LogscaleF_alpha}
\end{figure} 

Figure~\ref{fig:LogscaleF_alpha} presents Fig.~2 in the main text, but with a logarithmic ordinate. According to simulations, for the naive information engine the critical feedback gain 
$\astar = 0$ is the only solution below a critical signal-to-noise ratio $\snrc$, as shown in Fig.~\ref{fig:LogscaleF_alpha}(a). At low SNR, experimental determinations of the critical feedback gain $\astar$ do not agree with numerical simulations (Fig.~\ref{fig:LogscaleF_alpha}(a)), denoted as red hollow markers. Although experiments find $\astar > 0$, the naive output powers $\dot{F}$ are still an order of magnitude smaller than those of the Bayesian information engine (Fig.~\ref{fig:LogscaleF_alpha}(b)). Figure~\ref{fig:LogscaleF_alpha}(c) shows the corresponding input trap powers $\dot{W}$, which remain small at low SNR.

\section{Theoretical arguments for the phase transition in the naive information engine}

Empirically, the naive estimator can extract free energy only above a critical signal-to-noise ratio $\snrc~\approx~0.7$.

We define the critical SNR as the highest value for which $\astar = 0$ is the only solution for which $\dot W = 0$. 
Note that $\alpha=0$ corresponds to never moving the trap and therefore not doing any work on the bead. 
When no feedback is performed, the stationary distribution of the bead inside the trap is the equilibrium Boltzmann distribution,
\begin{align}
    p(x_k|\lambda_k) = \mathcal{N}(x_k;\lambda_k-\dg,\, 1)\,. \label{eq:equilibrium_dist}
\end{align}
This equation, the bead's propagator $p_x(x_{k+1}|x_k,\lambda_k)$ in Eq.~\eqref{eq:OU_dynamics}, and the measurement distribution $p_y(y_k|x_k)$ in Eq.~\eqref{eq:meas_dist} together allow us to evaluate the input work,
\begin{widetext}
\begin{subequations}
\begin{align}
    \left\langle \Delta W_{k+1}\right\rangle &= \frac{1}{2}\left\langle \left(x_{k+1} - \lambda_{k+1} \right)^2 - \left(x_{k+1} - \lambda_k \right)^2 \right\rangle \label{eq:work_naive_1}\\
    &= \frac{1}{2}\left\langle \left[x_{k+1} - \lambda_{k} - \Theta(y_k-\lambda_k) \, \alpha (y_k-\lambda_k) \right]^2 - \left(x_{k+1} - \lambda_k \right)^2 \right\rangle \label{eq:work_naive_2}\\
    &= \frac{1}{2} \int \dd{\lambda_k} p(\lambda_k)\int \dd{x_{k+1}} \int \dd{ x_k} \int\limits_{\lambda_k}^\infty \dd{y} \, p_x(x_{k+1}|x_k,\lambda_k) \, p_y(y_k|x_k) \, p(x_k|\lambda_k) \nonumber \\
    &\hspace{6cm}\times\left\{ \left[ x_{k+1}\!-\!\lambda_k - \alpha \left(y_k\!-\!\lambda_k \right) \right]^2 - \left(x_{k+1}\!-\!\lambda_k \right)^2 \right\} \label{eq:work_naive_3}\\
    &= \frac{1}{2} \int \dtot\lambda_k\, p(\lambda_k)\int \dtot\tilde x^+\int \dtot \tilde x\int\limits_{0}^\infty \dtot\tilde y \, p_x(\tilde x^+\!+\! \lambda_k|\tilde x + \lambda_k,\lambda_k) \, p_y(\tilde y + \lambda_k|\tilde x + \lambda_k) \, p(\tilde x + \lambda_k|\lambda_k)\nonumber\\
    &\hspace{10cm}\times\left[ \left( \tilde x^+ - \alpha\tilde y \right)^2 - \left(\tilde x^+\right)^2 \right] \label{eq:work_naive_4}\\
    &= \frac{1}{2} \int \dtot\lambda_k\, p(\lambda_k)\int \dtot \tilde x^+\int \dtot \tilde x\int\limits_{0}^\infty \dtot\tilde y \, \mathcal{N}\left[\tilde x^+;\tilde x \e^{-\ts} - \dg (1-\e^{-\ts}), \, 1-\e^{-2\ts} \right] \, \mathcal{N}(\tilde y; \tilde x,\, \sigma_\mrm{m}^2) \, \mathcal{N}(\tilde x;-\dg,1)\nonumber\\
    &\hspace{10cm}\times\left[ \left( \tilde x^+ - \alpha\tilde y \right)^2 - \left(\tilde x^+\right)^2 \right] \label{eq:work_naive_5}\\
    &= \frac{\alpha}{4}\left[ \alpha\left( \dg^2 + 1 + \sigma_\mrm{m}^2 \right) - 2\dg^2 - 2\e^{-\ts} \right]\left[ 1- \erf\left(\frac{\dg}{\sqrt{2(1+\sigma_\mrm{m}^2)}}\right) \right] + \sqrt{\frac{1+\sigma_\mrm{m}^2}{8\pi}} \, \alpha (2-\alpha) \dg \exp\left[-\frac{\dg^2}{2(1+\sigma_\mrm{m}^2)}\right]\,, \label{eq:work_naive_end}
\end{align}
\end{subequations}
where we substituted $\tilde x = x_k-\lambda_k$, $\tilde y = y_k - \lambda_k$, and $\tilde x^+ = x_{k+1} - \lambda_k$ in line~\eqref{eq:work_naive_4}, and used the normalization of the unknown $p(\lambda_k)$ in line~\eqref{eq:work_naive_end}.

Solving $\left\langle \Delta W_{k+1}\right\rangle =0$ for $\alpha$ yields solutions $\astar=0$, and
\begin{align}
    \astar &= 2\frac{\sqrt{2(1+\sigma_\mrm{m}^2)}\, \dg \exp\left[-\frac{\dg^2}{2(1+\sigma_\mrm{m}^2)} \right] - \sqrt{\pi}\left(\dg^2+\e^{-\ts}\right)\left[1-\erf \frac{\dg}{\sqrt{2(1+\sigma_\mrm{m}^2)}} \right] }{\sqrt{2(1+\sigma_\mrm{m}^2)}\, \dg \exp\left[-\frac{\dg^2}{2(1+\sigma_\mrm{m}^2)} \right] - \sqrt{\pi}\left(\dg^2+1+\sigma_\mrm{m}^2\right)\left[1-\erf 
    \frac{\dg}{\sqrt{2(1+\sigma_\mrm{m}^2)}} \right]}\,. \label{eq:solution_alpha}
\end{align}
\end{widetext}
At the critical $\sm$, the solution $\astar$ bifurcates. We determine the critical SNR from Eq.~\eqref{eq:solution_alpha} by solving
\begin{align}
    \astar = 0
\end{align}
for $\sm$, which can only be done numerically. For $\ts=1/41$ and $\dg=0.8$, the parameters used in Figs.~2 and 3,
\begin{align}\label{eq:critical_alpha_theory}
    \snrc = \left(\sigma_\mrm{m}^{\mrm{c}}\right)^{-1} \approx 0.64\,.
\end{align}
Figure~2 in the main text shows that this critical SNR indeed empirically coincides with
the emergence of a positive $\astar$. We note that the accuracy of this theoretical estimate (and thus the numerical precision we give) is limited by the accuracy of experimental parameters such as $\delta_g$ and $\ts$ which depend on the bead size.

\section{Experimental evidence supporting a phase transition in the naive information engine}

\begin{figure}[t]  
    \centering
    \includegraphics[clip, width=1.0\linewidth]{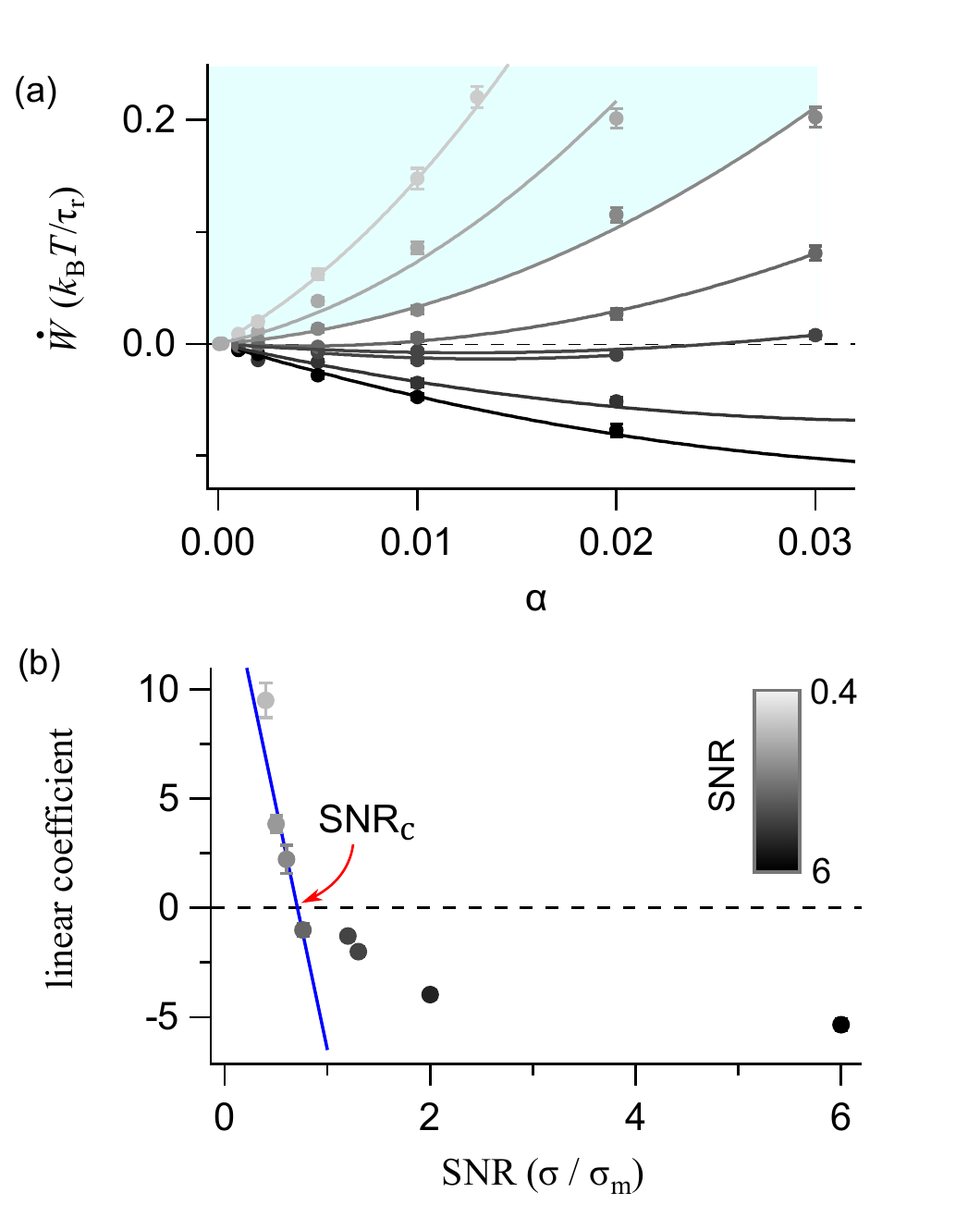}
    \caption{Phase transition in naive information engine. (a) Trap power as a function of feedback gain $\alpha$ for different signal-to-noise ratios: 0.4, 0.5, 0.6, 0.8, 1.2, 1.3, 2, and 6 (different shades of gray, scale bar in (b)). Markers: experiment; curves: fit to quadratic functions. Blue shaded region: SNR values for which $\dot{W}>0$ for all values of $\alpha$. (b) Linear coefficients from quadratic fits (curves in (a)), for different SNRs. Blue line: linear fit to first four points. Error bars each denote standard error of the mean in (a) and parameter fit error in (b).}
    \label{fig:naive_W_alpha}
\end{figure}

In this section, we present detailed experimental evidence for a phase transition in the naive information engine. Figure~\ref{fig:naive_W_alpha} (a) shows the trap power as a function of the feedback gain $\alpha$ for different SNR, where the markers in the blue shaded region represent SNR values for which only positive trap powers ($\dot{W}>0$) were measured. The experimental points are fit to a quadratic function; Fig.~\ref{fig:naive_W_alpha} (b) shows the linear coefficients of the fits in Fig.~\ref{fig:naive_W_alpha} (a) as a function of SNR. The measured slope switches from negative at high $\snr \gtrsim 0.8$ to positive at $\snr \lesssim 0.6$, consistent with $\snrc$ determined in Eq.~\eqref{eq:critical_alpha_theory}. To measure the critical signal-to-noise ratio (where the linear coefficient is zero), the linear coefficients for SNRs 0.4--0.8 are linearly fit, giving an estimated zero linear coefficient and hence $\snrc$ at $0.7\pm0.1$. In the naive information engine experiments, for $\snr<\snrc$, we set $\alpha>\astar$ because we cannot find an $\alpha$ that leads to $\dot{W}<0$. Thus, the $\astar$ we report in such cases (red hollow markers in Fig.~3, main text) is an upper bound on the
$\astar$ that satisfies $\dot{W}\gtrsim 0$: Either there is a smaller positive $\astar$ that we could not detect experimentally because differences in $\dot{W}$ are too close to zero, or the only value that (always) enforces $\dot{W}=0$ is $\astar=0$. The corresponding $\dot{F}$ measured using the upper-bound value of $\astar$ then is an upper bound on the output power for $\snr<\snrc$.

Importantly, the experimental trap powers in Fig.~\ref{fig:naive_W_alpha} need not match the work in Eq.~\eqref{eq:work_naive_end} because that equation assumed that the bead's stationary-state position distribution is the equilibrium distribution~\eqref{eq:equilibrium_dist}, which only strictly holds when $\alpha=0$.

\section{The Bayesian information engine has no phase transition}
Here, we present theoretical arguments to support the claim that there is no corresponding ``phase transition'' for the information engine that estimates the bead position using filtered measurements.

\subsection{Feedback gain}
\begin{figure}[htbp!]
    \centering
    \includegraphics[clip, width=0.8\linewidth]{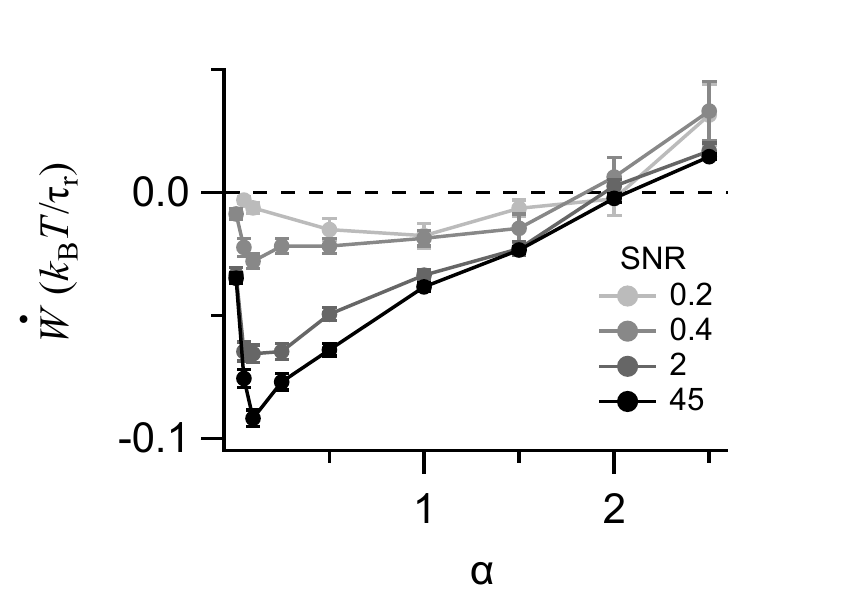}
    \caption{Experiments showing the Bayesian information engine has no phase transition. Trap power $\dot{W}$ as a function of feedback gain $\alpha$ for different signal-to-noise ratios (different shades of gray). Solid curves connect solid markers to ease visualization. Error bars each denote the standard error of the mean.}
    \label{fig:filter_work_alpha}
\end{figure}

First, we show that feedback gain $\alpha=2$ generally produces vanishing input work. Using the predictive Kalman filter [Eq.~(6)] in the feedback rule [Eq.~(5)] with $\alpha=2$, the trap work is
\begin{widetext}
\begin{subequations}
\begin{align}
    \left\langle \Delta W_{k+1}\right\rangle &= \frac{1}{2}\left\langle \left(x_{k+1} - \lambda_{k+1} \right)^2 - \left(x_{k+1} - \lambda_k \right)^2 \right\rangle \label{eq:work_filter_1}\\
    &= \frac{1}{2}\left\langle \left[ x_{k+1} - \lambda_{k} - 2\, \Theta(\hat x_{k+1}-\lambda_k) \, (\hat x_{k+1}-\lambda_k) \right]^2 - \left(x_{k+1} - \lambda_k \right)^2 \right\rangle \label{eq:work_filter_2}\\
    &= \frac{1}{2} \int \dtot\lambda_k\, p(\lambda_k)\int \dtot x_{k+1}\int\limits_{\lambda_k}^\infty \dtot \hat x_{k+1} \, p(x_{k+1},\hat x_{k+1}|\lambda_k) \left\{ \left[ x_{k+1}\!-\!\lambda_k - 2 \left(\hat x_{k+1}\!-\!\lambda_k \right) \right]^2 - \left(x_{k+1}\!-\!\lambda_k \right)^2 \right\} \label{eq:work_filter_3}\\
    &= 2 \int \dtot\lambda_k\, p(\lambda_k)\int\limits_{\lambda_k}^\infty \dtot \hat x_{k+1} \, p(\hat x_{k+1}|\lambda_k)\,\left( \hat x_{k+1} - \lambda_k \right)  \underbrace{\int \dd{x_{k+1}} \, p(x_{k+1}|\hat x_{k+1},\lambda_k)\,\left( x_{k+1} - \hat x_{k+1} \right)}_{=\,0}  \label{eq:work_filter_4}\\
    &= 0\,,
\end{align}
\end{subequations}
\end{widetext}
where \eqref{eq:work_filter_4} used the fact that the posterior distribution $p(x_{k+1}|\hat x_{k+1},\lambda_k)$ of the true bead position $x_{k+1}$ is symmetric around the filter estimate $\hat x_{k+1}$, which in this case is the mean of the Gaussian posterior. Figure~\ref{fig:filter_work_alpha} shows the input work as a function of the feedback gain $\alpha$ for different signal-to-noise ratios. $\alpha=2$ makes the input work vanish.

\subsection{Free-energy gain}

With feedback gain $\alpha=2$ and the feedback rule in Eq.~(5), the free energy-gain per time step, Eq.~(7), becomes
\begin{subequations}
\begin{align}
    &\left\langle\Delta F_{k+1} \right\rangle = \dg \left\langle \lambda_{k+1}-\lambda_k \right\rangle\\
    &\quad = 2\dg \int \dtot\lambda_k\, p(\lambda_k)\underbrace{\int\limits_{\lambda_k}^\infty \dtot\hat x_{k+1}\, p(\hat x_{k+1}|\lambda_k)\, \left( \hat x_{k+1}\!-\! \lambda_k \right)}_{> \, 0} \label{eq:filter_free_energy_2}\\
    &\quad >0\,,
\end{align}
\end{subequations}
where we used the fact that the probability density $p(\hat x_{k+1}|\lambda_k)$ is positive and always has a tail (however small) reaching into the region $\hat x_{k+1} > \lambda_k$. Hence, there is no phase transition in the Bayesian information engine.
Note that for the naive engine 
\begin{align}
    \langle\Delta F_{k+1}\rangle~=~\alpha \dg\int\mrm d\lambda_k\,p(\lambda_k) \int_{\lambda_k}^\infty\mrm d y_k\,p(y_k|\lambda_k)(y_k-\lambda_k)\,,
\end{align}
which vanishes for $\alpha=0$.

To corroborate the finding that the Bayesian information engine has no phase transition, we calculate the low-SNR limit of the free-energy gain per time step. When SNR is low, feedback is applied rarely and hence the stationary distribution of bead positions approximates the equilibrium Boltzmann distribution, $p(x_{k+1}|\lambda_k) = \mathcal{N}(x_{k+1};\lambda_k-\dg,1)$. Hence, the distribution $\hat p(\hat x_{k+1}|\lambda_k)$ of filter estimates is also  approximately Gaussian,
\begin{align}
    \hat p (\hat x_{k+1}|\lambda_k) = \mathcal{N}\left( \hat x_{k+1};\hat \mu, \, \hat c \right)\,. \label{eq:filter_dist_Ansatz}
\end{align}
The mean and variance~\eqref{eq:KF_predict} of the predictive Kalman filter dictate the steady-state the posterior distribution, of the true bead position $x_{k+1}$ given the filter estimate $\hat x_{k+1}$\,,
\begin{align}
    p(x_{k+1}|\hat{x}_{k+1}) = \mathcal{N}\left(x_{k+1};\hat x_{k+1},\, \hat P^\mathrm{SS}\right)\,,
\end{align}
allowing us to solve for $\hat p(\hat x_{k+1}|\lambda_k)$,
\begin{align}
    \hat p(\hat x_{k+1}|\lambda_k) = \int \dd{u} \, p(x_{k+1}|\hat x_{k+1}=u) \, \hat p(u|\lambda_k)\,, \label{eq:stationary_filter_def}
\end{align}
such that
\begin{subequations}
\begin{align}
    \hat\mu &= \lambda_k-\dg\\
    \hat c &= 1 - \hat P^\mathrm{ss}\,,
\end{align}
\end{subequations}
are the mean and variance of the filter state in Eq.~\eqref{eq:stationary_filter_def}.
 
Inserting these cumulants into Eq.~\eqref{eq:filter_dist_Ansatz} and evaluating Eq.~\eqref{eq:filter_free_energy_2} again using the normalization of $p(\lambda_k)$ gives

\begin{widetext}

\begin{align}
     \left\langle\Delta F_{k+1} \right\rangle &= \dg \sqrt{\frac{2(1-\hat P^\mathrm{ss})}{\pi}} \, \exp\left[-\frac{\dg^2}{2(1-\hat P^\mathrm{ss})}\right] - \dg^2\left[ 1-\erf
     \frac{\dg}{\sqrt{2(1-\hat P^\mathrm{ss})}} 
     \right]\,.
\end{align}
Expanding the stationary filter variance~\eqref{eq:steady_filter_var_sol} for small $\snr = 1/\ssm$ yields
\begin{align}
    \hat P^\mathrm{ss} &= 1 - \frac{\mathrm{SNR}^2}{\e^{2\ts}-1} + \mathcal{O}(\snr^4)
\end{align}
Then, with $\erf(x) \approx 1- \e^{-x^2}\left(1/x - 1/2x^3 \right)/\sqrt{\pi}$ for $x\to \infty$,
\begin{align} 
    \left\langle\Delta F_{k+1} \right\rangle &\approx \sqrt{\frac{2}{\pi}} \left(\frac{\snr}{\sqrt{\e^{2\ts}-1}}\right)^3 \exp\left[- \frac{\dg^2 (\e^{2\ts}-1)}{2\,\snr^2}\right] > 0\,,
\end{align}
illustrating that the Bayesian information ratchet extracts free energy at all signal-to-noise ratios.

\end{widetext}
\providecommand{\noopsort}[1]{}\providecommand{\singleletter}[1]{#1}%

\end{document}